\documentclass[sigconf]{acmart}
\setcopyright{none} 
\renewcommand\footnotetextcopyrightpermission[1]{} 
\acmConference[]{}{}{}

\usepackage{float}
\usepackage{booktabs}
\usepackage{tablefootnote}
\usepackage{flushend}

\newcommand{\mt}{\textsc{Sesame}}  

\AtBeginDocument{%
  \providecommand\BibTeX{{%
    \normalfont B\kern-0.5em{\scshape i\kern-0.25em b}\kern-0.8em\TeX}}}

\begin{document}

\title{Semistructured Merge with Language-Specific Syntactic Separators}


\author{Guilherme Cavalcanti}
\affiliation{%
  \institution{Federal Institute of Pernambuco}
  \city{Belo Jardim}
  \country{Brazil}
}

\author{Paulo Borba}
\author{Leonardo dos Anjos}
\author{Jonatas Clementino}
\affiliation{%
  \institution{Informatics Center, \\ Federal University of Pernambuco}
  \city{Recife}
  \country{Brazil}
}


\renewcommand{\shortauthors}{Cavalcanti et al.}

\begin{abstract}

Structured merge tools exploit programming language syntactic structure to enhance merge accuracy by reducing spurious conflicts reported by unstructured tools. 
By creating and handling full ASTs, structured tools are language-specific and harder to implement. 
They can also be computationally expensive when merging large files.
To reduce these drawbacks, semistructured merge tools work with partial ASTs that use strings to represent lower level syntactic structures such as method bodies, and rely on unstructured tools to merge them.  
This, however, results in merge accuracy loss. 
To improve accuracy without compromising semistructured merge benefits, we propose a tool that leverages language-specific syntactic separators to infer structure without parsing.
We still resort to an unstructured tool to merge lower level structures, but only after preprocessing the code so that text in between separators such as curly braces appear in separate lines.  
This way we emulate the capabilities of structured merge tools while avoiding their drawbacks. 
By comparing our tool with a robust implementation of semistructured merge, we find that our tool substantially reduces the number of spurious conflicts.
We also observe significant but less substantial reductions on the overall number of reported conflicts, and of files with conflicts.
However, similar to structured tools, our tool lets more merge conflicts go undetected.
Our tool shows significant improvements over unstructured tools widely used in practice.
Finally we observe that exploiting language-specific syntactic separators introduces unique textual alignment challenges. 
\end{abstract}

\ccsdesc[500]{Software and its engineering~Software maintenance tools}
\ccsdesc[500]{Software and its engineering~Software configuration management and version control systems}

\keywords{Semistructured merge; Merge conflicts; Configuration management; Software evolution}



\maketitle

\section{Introduction}


Unstructured merge tools have been in use for over 40 years~\cite{mens2002state}, and are essential for team software development. 
These tools rely solely on textual analysis to detect and resolve merge conflicts~\cite{Khanna2007}. 
They are fast and can handle any type of textual content, including source code in any language. 
However, they often report spurious merge conflicts, and fail to report conflicts that could lead to compilation and execution issues.
For example, unstructured merge reports a conflict when asked to integrate rearrangeable declarations (such as method and field declarations) separately added by two developers to the same area of the source code~\cite{apel2012structured}. 
Simply juxtaposing them in any order would be a correct merge, but an unstructured tool is not able to do that as it recognizes program lines instead of declarations.

To improve merge accuracy, researchers have proposed tools that consider the structure of the source code being merged~\cite{westfechtelStructured,judithcdiff,binkley1995program,buffenbarger1995,apiwattanapong2007jdiff,apel2012structured,LeBenich:2017,Zhu:2018,shen2019intellimerge,zhu2022mastery,larsen2022spork}. 
Such structured merge tools go beyond simple textual, line-based, analysis by parsing the code and handling abstract syntax trees (ASTs).
They employ tree matching and combination algorithms. 
Because of that, they are language-specific and demand significant implementation effort for each language they handle.
They can also be computationally expensive when merging files with many or large declarations~\cite{LeBenich:2017}. 

To reduce these drawbacks, researchers proposed semistructured merge tools~\cite{apel2011semistructured, cavalcantietal2017,tavares2019semistructured} 
that attempt to hit a sweet spot between unstructured and structured merge by partially exploring the underlying programming language syntactic structure.
The idea is to work with partial ASTs that use strings to represent lower level syntactic structures such as method bodies, statements, and expressions.
To merge these parts of the code, semistructured tools simply rely on textual, line-based, analysis by invoking an unstructured tool.
Although bringing significant benefits in comparison to unstructured merge, this, however, results in merge accuracy loss comparing with structured merge~\cite{cavalcanti2019impact,seibt2021leveraging}. 
 

To improve accuracy without compromising semistructured merge benefits, here we propose \mt, a semistructured merge tool that leverages language-specific syntactic separators (like `\texttt{\{}' and `\texttt{\}}' in Java) to infer structure without parsing. 
We still resort to an unstructured tool to merge lower level structures, but only after preprocessing the code so that text in between separators appear in separate lines. 
This way we more closely emulate the capabilities of structured merge tools while avoiding their drawbacks.

Whereas structured tools rely on the occurrence of specific token sequences to determine which AST node to create and latter match, our tool relies on separators to delimit areas of the code to match for textual merge. 
Thus, on lower level structures, \mt\ actually matches sequences of characters delimited not only by `\texttt{\textbackslash{n}}' (as in unstructured, line-based tools) but also by language-specific separators. 
Line-based tools solely rely on line breaks as the context delimiter for conflict detection, reporting conflicts whenever one tries to integrate changes made by two developers to the same or consecutive lines. 
In contrast, \mt\ only reports conflicts when one tries to integrate changes made to the same text area delimited by the separators.

We evaluate \mt\ by comparing it with a robust implementation of standard semistructured merge, on a dataset of merge scenarios--- quadruples $(\mathit{base}, \mathit{left}, \mathit{right}, \mathit{merge})$, where $\mathit{left}$ and $\mathit{right}$ are the parent commits integrated by $\mathit{merge}$--- from open-source projects gathered from GitHub.  
We replay the merge operation with the two tools on those scenarios to collect evidence on how often the tools report different results, merge conflicts, files with merge conflicts, false positives (conflicts incorrectly reported by a merge tool), and false negatives (actual conflicts missed by a merge tool). 
To also assess the benefits of our tool in relation to the standard practice, we also compare it with an unstructured merge tool.
In particular, we answer the following research questions:
\begin{itemize}
    \item \textbf{RQ1}: Does \mt\ reduce the number of merge conflicts and files with merge conflicts compared to standard semistructured and unstructured merge tools?
    \item \textbf{RQ2}: Does \mt\ reduce the number of merge conflicts identification errors compared to standard semistructured and unstructured merge tools?
\end{itemize}

Our results show that \mt\ outperforms standard semistructured and unstructured merge in terms of conflict reduction; our tool reduces the number of merge conflicts by 41\% and 13\%, respectively. 
We observe a similar pattern in favour of our tool when considering the number of files with merge conflicts.
Additionally our tool is far better in avoiding spurious conflicts (false positives); it substantially reduces them in comparison to unstructured and semistructured merge (91\% and 88\%, respectively). 
However, similar to structured tools and conforming to what is observed in previous studies~\cite{cavalcantietal2017,cavalcanti2019impact,tavares2019semistructured,seibt2021leveraging}, the increased ability to resolve merge conflicts comes at the price of missing actual, proper conflicts. 
\mt\ is no different: we found no evidence that \mt\ misses fewer actual conflicts (false negatives) than unstructured and standard semistructured merge. 
Based on these findings, \mt\ makes a step forward on dealing with the merge inaccuracy of the state-of-practice textual merge tools without the implementation effort and runtime overhead of structured merge tools. 
We find evidence that our tool shows significant improvements over unstructured tools widely used in practice.
Finally we observe that exploiting language-specific syntactic separators introduces unique
textual alignment challenges. 

\section{Motivation}
\label{sec:motivation}



\begin{figure}[t!]
  \includegraphics[width=.85\linewidth]{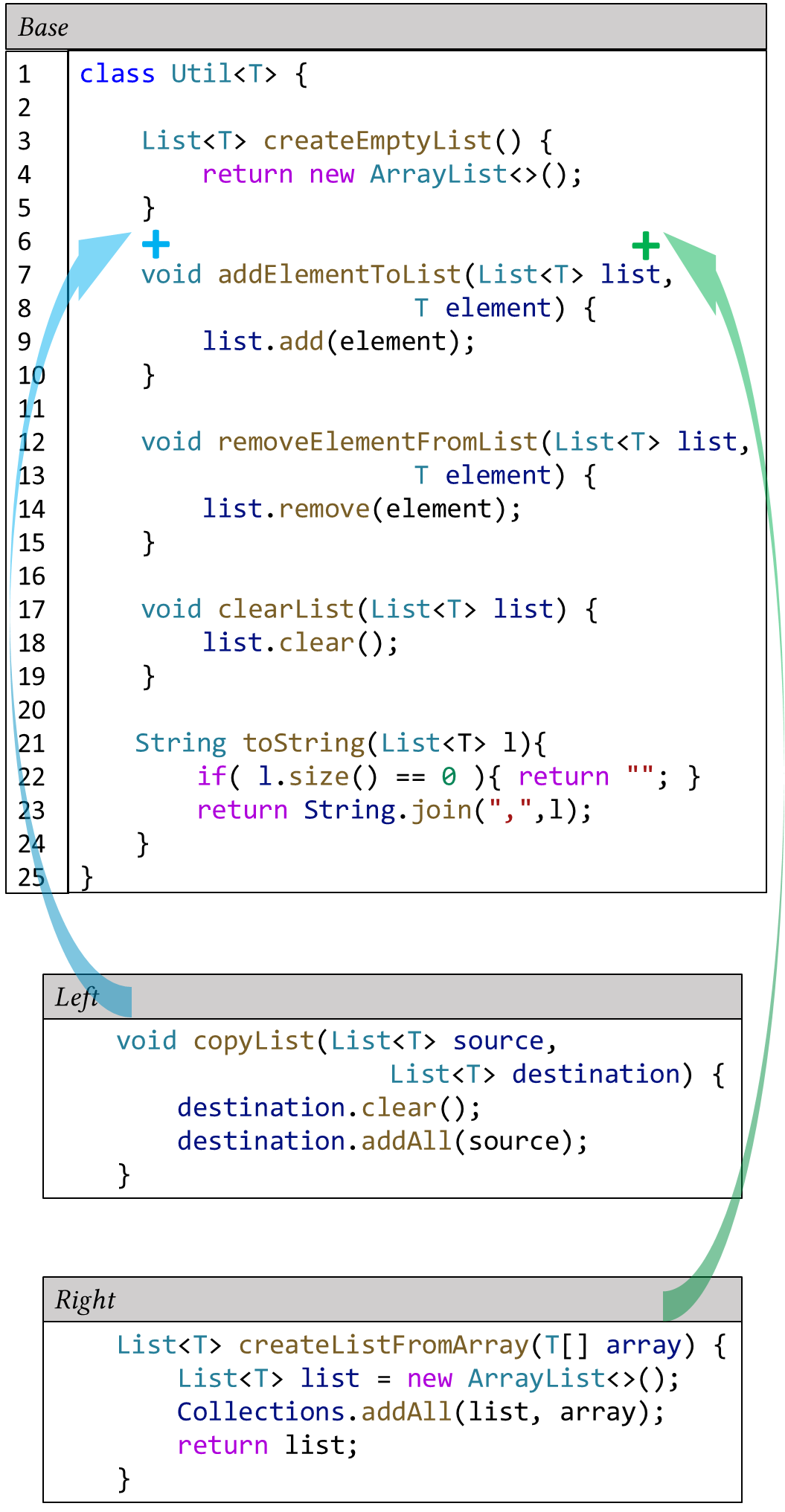}
  \caption{Independent changes to the same \textit{Util} class. 
  }
  \label{fig:util_changes}
\end{figure}

To better explain the differences between the kinds of merge tools we discuss in the previous section, consider the example in Figure~\ref{fig:util_changes}. 
On the top it shows the base version of the \textit{Util} Java class. 
Consider that two developers, \emph{Left} and \emph{Right}, independently introduce declarations to that class. 
The first developer, adds the \textit{copyList} method declaration right before the \textit{addElementToList} declaration. 
Unaware of \emph{Left}'s modification, \emph{Right} adds the \textit{createListFromArray} method declaration, at the same text area, but in its local copy of the file. 
Subsequently, the task at hand is to merge these disparate local file versions into a unified version of the \textit{Util} class. 
The goal is to synthesize a consolidated version of the class that includes the changes introduced by both developers.

\begin{figure}[h]
  \includegraphics[width=.85\linewidth]{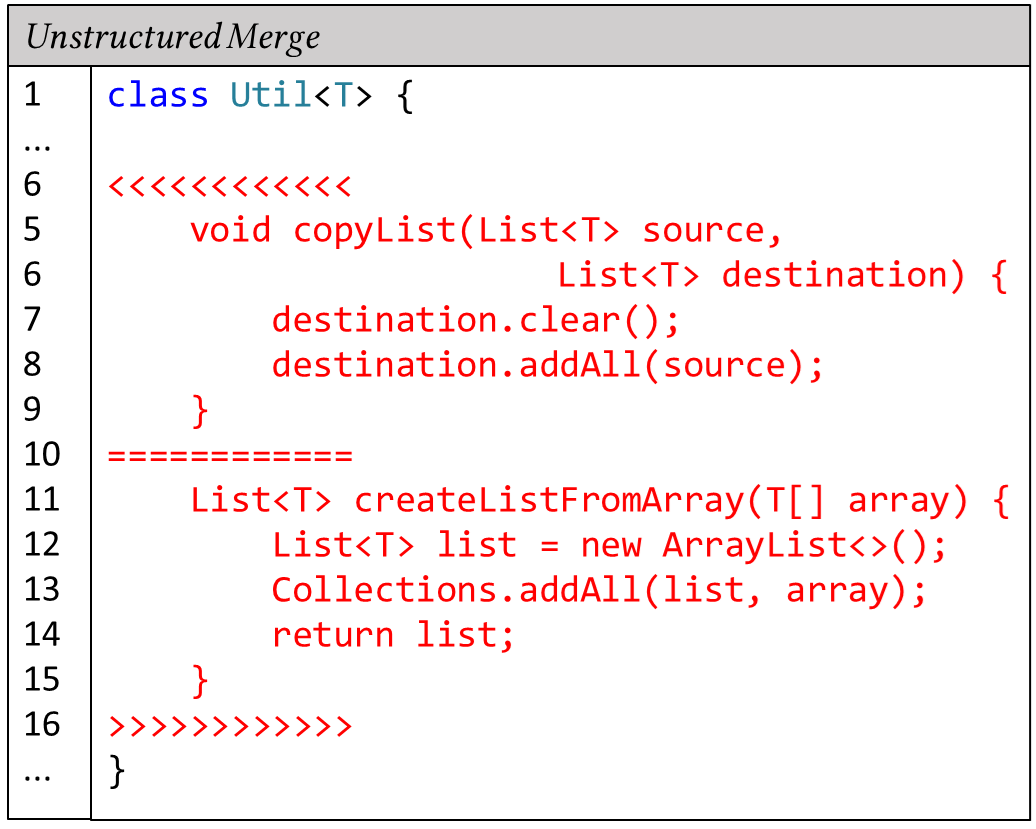}
  \caption{Unstructured merge of different versions of the same \textit{Util} class resulted in merge conflict.}
  \label{fig:util_merge}
\end{figure}

An unstructured merge tool contrasts the two modified files, line by line, relative to their shared common ancestor, identifying distinct sequences of lines known as ``chunk''. 
The tool then verifies whether the three file versions (\emph{base}, \emph{left}, \emph{right}) share a common text region that separates the contents of the modified chunks. 
If such a separator is not found, the tool reports a merge conflict. 
In our example, an unstructured merge tool reports the merge conflict illustrated in Figure~\ref{fig:util_merge}.
As both developers change line 6 
of the \emph{Base} version of the file, there is no common separator, so unstructured merge is unable to integrate the changes.

\begin{figure}[h!]
  \includegraphics[width=.85\linewidth]{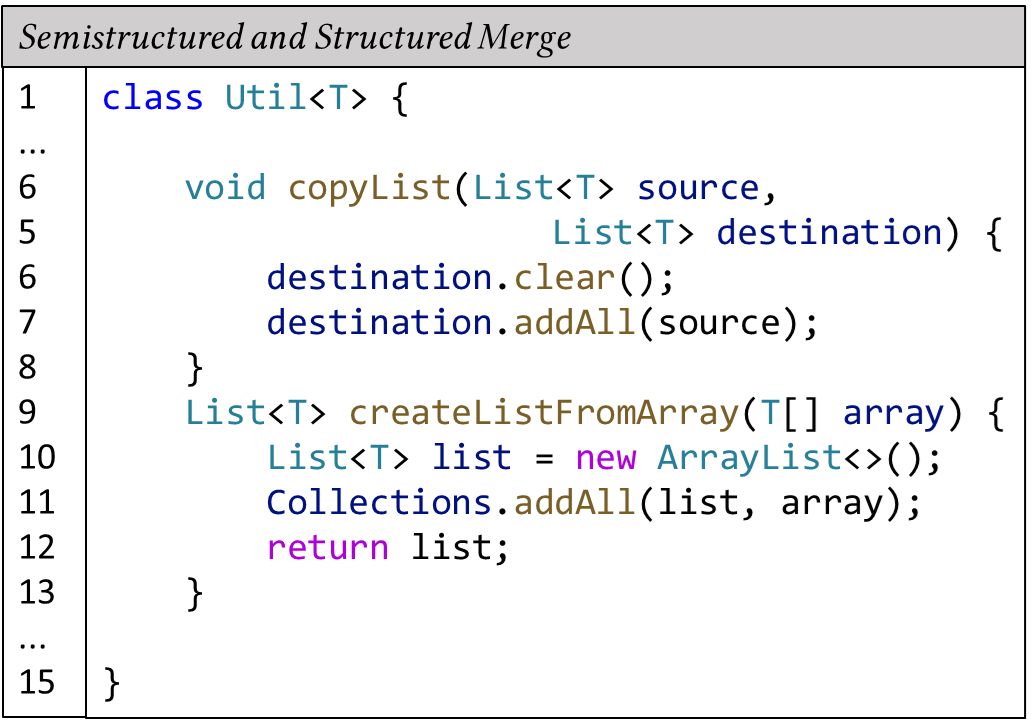}
  \caption{Structured and semistructured successful merge of different versions of the same \textit{Util} class.}
  \label{fig:util_structuredmerge}
\end{figure}

Note, however, that both changes could be safely integrated by a merge tool aware that the changes correspond to unrelated Java method declarations, whose order does not matter. 
Such a tool could simply juxtapose the declarations, no matter in which order.
This motivated the development of syntax-based, structure aware merge tools, such as structured and semistructured merge tools~\cite{apel2011semistructured,apel2012structured,cavalcantietal2017,LeBenich:2017,Zhu:2018,shen2019intellimerge,zhu2022mastery,larsen2022spork}.
Such tools generate abstract syntax trees (ASTs) for each version of the files being integrated. These trees are then compared to identify common, introduced, or removed nodes in each tree. 
The tools then report merge conflicts when changes are related to the same tree node. 
As a result, semistructured and structured merge tools avoid spurious conflicts such as the one in Figure~\ref{fig:util_merge}. 
In that example, the changes occur on the same text area, but each method declaration is attributed to a distinct AST node. 
Consequently, semistructured and structured merge tools interpret that these modifications are unrelated, and correctly accept both contributions, yielding the class version in Figure~\ref{fig:util_structuredmerge}.

\begin{figure}[h]
  \includegraphics[width=.85\linewidth]{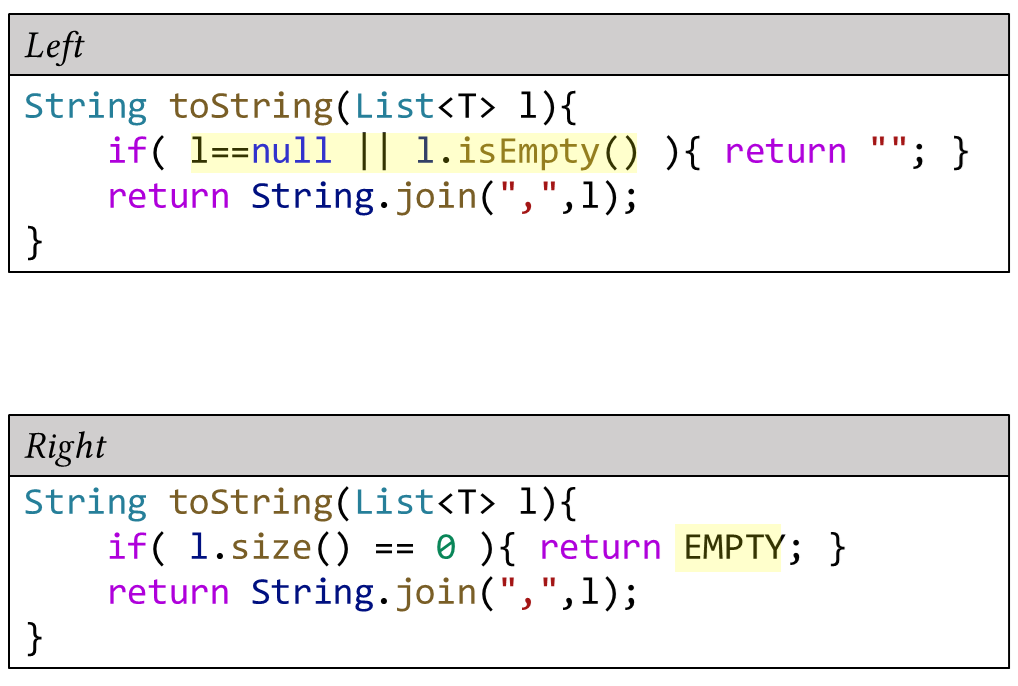} 
  \caption{Independent changes (highlighted in yellow) to the same \textit{toString} method. 
  }
  \label{fig:toString_changes}
\end{figure}

Now consider that besides adding the \textit{copyList} and \textit{createListFromArray} method declarations, the developers change the \textit{toString} method in different ways. 
As illustrated in Figure~\ref{fig:toString_changes}, this method receives a list as argument and returns a string. 
\emph{Left} changes the \texttt{if} statement condition to check whether the list is \texttt{null} or empty. 
Independently, \emph{Right} extracted a constant from the condition body, which happens to be in the same line of the condition changed by \emph{Left}.

\begin{figure}[h]
  \includegraphics[width=.9\linewidth]{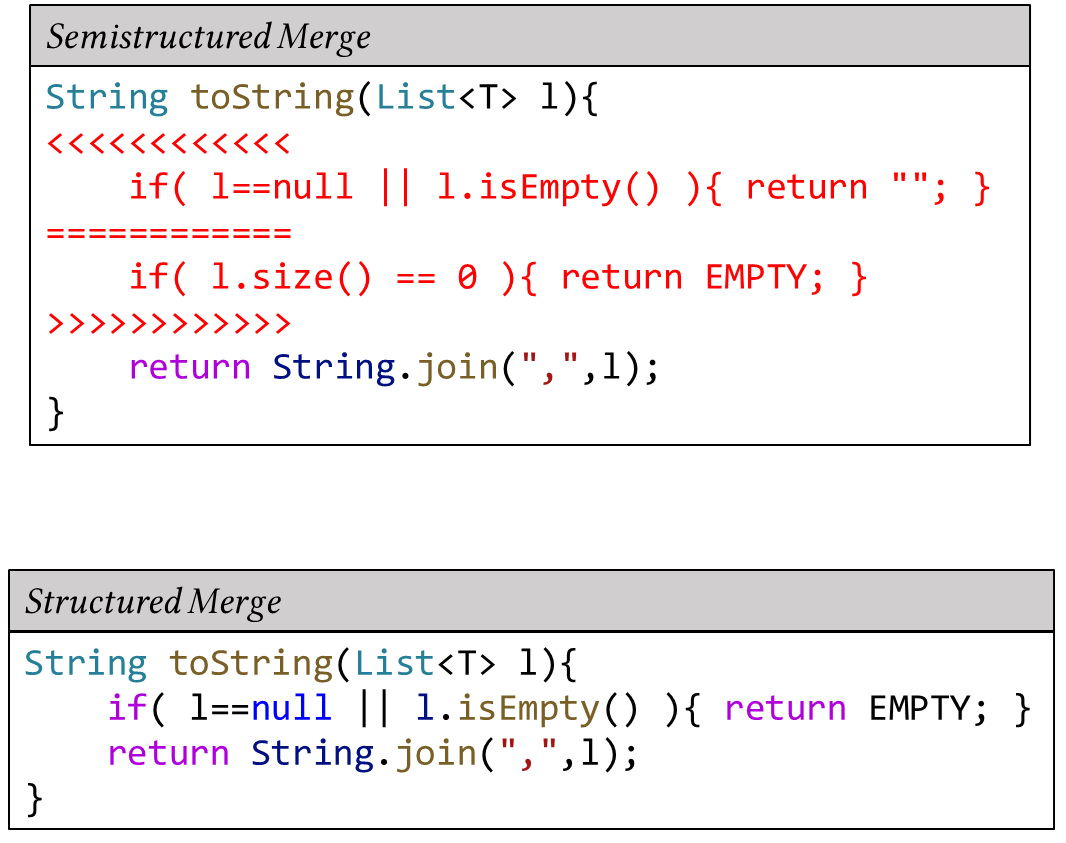}
  \caption{Structured and semistructured merge of different versions of the same \textit{toString} method.}
  \label{fig:toString_structuredmerge}
\end{figure}

A semistructured merge tool reports a conflict in this case because it invokes an unstructured tool to merge low level syntactic structures such as method bodies and statements. 
As the changes to be merged occur on the exact same line of code, we get the conflict shown in Figure~\ref{fig:toString_structuredmerge}. 
Once again this represents a spurious merge conflict. 
\emph{Right}'s change is a refactoring, thus being semantically neutral and free of any interference with the changes made by \emph{Left}. 
Semistructured merge, by only partially exploiting the syntactic structure of the code fails in this case.
Structured merge, however, represents method bodies as ASTs and reports no conflict in this case  because it detects that the changes occur in different AST nodes: one that represents the condition of the \texttt{if} statement, and the other that represents its body. 
  
\mt\ would yield the same result as structured merge in this case, as we will further discuss in the next section.
Although our tool does not represent statements as ASTs, avoiding the major drawbacks of structured tools, it exploits the `\texttt{\{}' separator that delimits the beginning of the \texttt{if} body.

\section{Semistructured Separator-based Merge}
\label{sec:solution}


Previous studies demonstrate that semistructured merge offers significant advantages over traditional unstructured merge for resolving merge conflicts~\cite{apel2011semistructured,cavalcantiassessing,cavalcantietal2017,cavalcanti2019impact,seibt2021leveraging}. 
However, by design, semistructured merge inherits some of the inaccuracies associated with unstructured merge when merging lower level syntactic structures such as statements and expressions, as illustrated in the previous section. 
This occurs because typical semistructured merge tools only create AST nodes up to class-level declarations in Java-like languages.
So such tools will have nodes for declarations such as imports, fields and methods, but these nodes contents are not represented as ASTs. 
The remaining programming language elements are simply represented as plain text, and are merged using a standard unstructured merge algorithm.

To avoid the merge inaccuracies resulting from that design decision, one could increase the level of structure and detail in the ASTs generated by semistructured merge, aligning it more closely with structured merge.
This, however, cannot be done without inheriting structured merge drawbacks of requiring an  excessively language dependent implementation that requires substantial per language effort.
Additionally, as the level of detail in ASTs increases, so does the computational costs  associated with matching these trees with expensive algorithms. 
Although such costs are likely manageable for most merge scenarios~\cite{apel2012structured,Lessenich:2015,LeBenich:2017}, it is a factor that should be taken into consideration for more complex merge scenarios.

To balance these two opposing design dimensions, and mitigate the \textit{inherited} inaccuracies of semistructured merge, we propose a technique that further explores the syntactic structure of source code while retaining the current level of detail of the generated ASTs of semistructured merge. 
We basically refine the unstructured merge process invoked by semistructured merge for merging low level syntactic structures.
Our idea is to consider language-specific separators as delimiters for matching and conflict detection. 
So, instead of relying solely on line matching, we propose a character sequence matching process that takes into account not only `\texttt{\textbackslash{n}}' as a separator, but also language-specific syntactic delimiters such as \texttt{‘\{’} and \texttt{‘\}’} in Java.

Line-based tools solely rely on line breaks as the context delimiter for conflict detection reporting conflicts whenever one tries to integrate changes made by two developers to the same or consecutive lines of the program text. 
In contrast, our technique only reports conflicts when one tries to integrate changes made to the same text area delimited by the separators.
By doing so, we effectively avoid the false merge conflict illustrated in Figure~\ref{fig:toString_structuredmerge}. 
Our technique considers `\texttt{\{}' as a delimiter for text areas, thus recognizing that the  changes to be integrated occur in distinct regions: before and after the `\texttt{\{}' in the second line of the \textit{toString} method declaration. 
The result is analogous to a full structured merge in that case, but without its associated drawbacks.

We implement this technique in the \mt\ tool, which builds upon both the widely used unstructured merge tool \texttt{diff3} and the robust semistructured merge tool \texttt{s3m}~\cite{cavalcantietal2017}, which is specific for merging Java code. 
We still resort to the unstructured tool to merge lower level structures, but only after preprocessing the code so that text in between separators appear in separate lines.
It works as follows:
\begin{enumerate}
    \item It inserts a new line before and after each language-specific separator (e.g. `\texttt{\{}' and `\texttt{\}}') in the different versions (\emph{base}, \emph{left}, \emph{right}) of the lower level syntactic structures to be merged.
    \item It marks the newly added lines with a distinctive placeholder (we adopt a stream of `\$' characters).
    \item It invokes unstructured merge on the marked versions of the lower level syntactic structures.
    \item It removes the new lines and placeholders from the result of unstructured merge.
\end{enumerate}

The list of separators can be easily customized for each programming language, or even to experiment with variations for the same language.

\begin{figure*}[t!]
  \includegraphics[width=.65\linewidth]{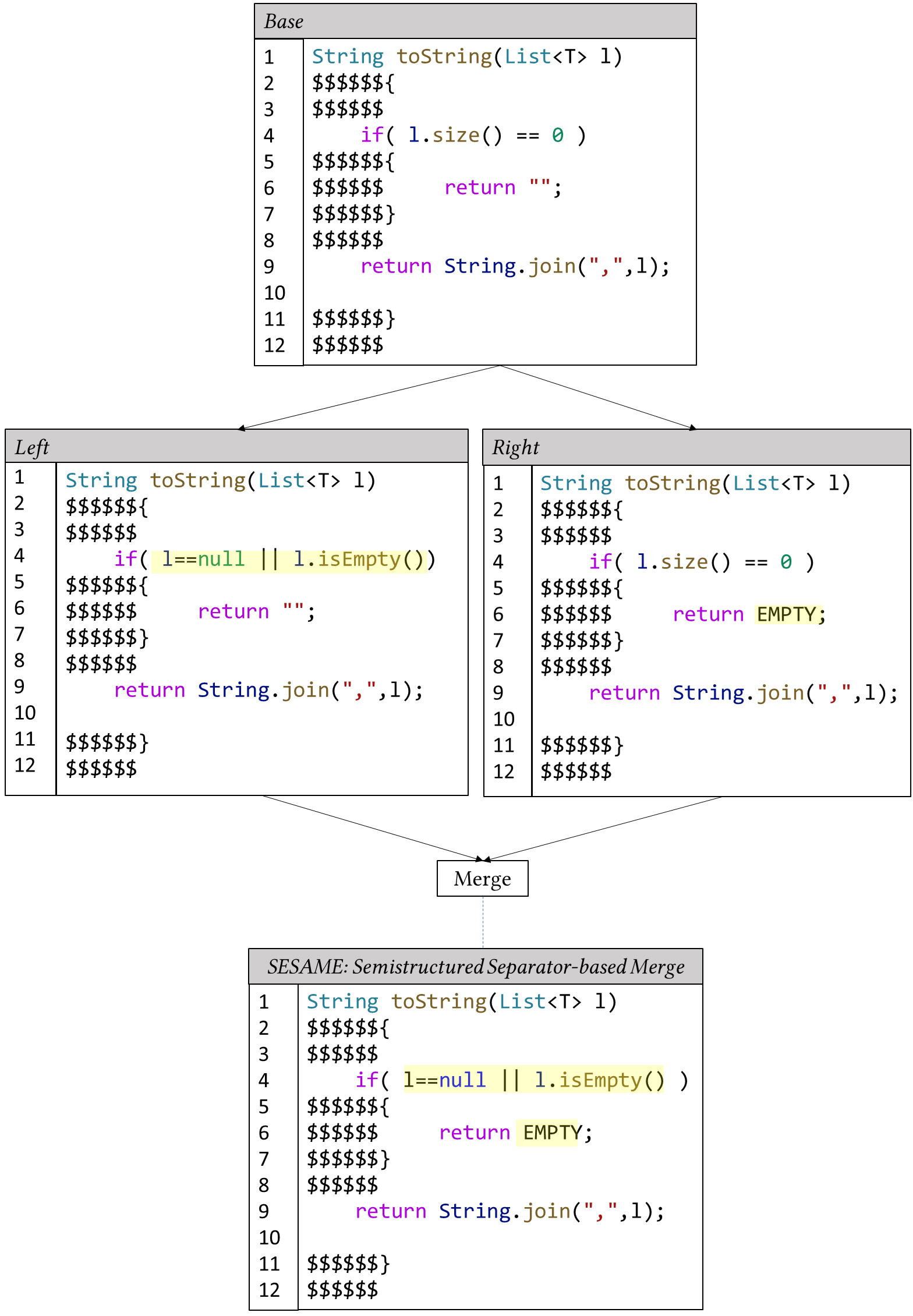}
  \caption{\textit{toString} method code transformation by \mt. Developers' changes highlighted in yellow. 
  }
  \label{fig:toString_transformation}
\end{figure*}

To illustrate this process, consider again the example in Figure~\ref{fig:toString_changes}. Consider that \mt\ is customized to use only curly braces (`\texttt{\{'} and \texttt{`\}}') as language-specific separators. 
In Java, these characters define the start and end of a code block, creating distinct text areas for analysis and matching. 
Initially, our tool inserts the new lines and placeholder both before and after every `\texttt{\{'} and \texttt{`\}}'.
This action is carried out individually for each file to be merged by the tool, generating a total of three temporary shadow files, each one corresponding to distinct versions of the original file: the base, left the right versions, shown in Figure~\ref{fig:toString_transformation}. 

Next \mt\ invokes unstructured merge on the three temporary files. 
The result of this step is then depicted in the bottom of Figure ~\ref{fig:toString_transformation}. 
Note that the merging process unfolds correctly, as evidenced by the successful accommodation of the changes, without merge conflicts. 
This success can be attributed to the fact that, now, the changes are placed in distinct text areas due to \mt\ preprocessing: while changes originating from the left version are located in line 4, those originating from the right version appear in line 6. 
This distribution of the changes across different and non-consecutive lines enables unstructured merge to properly merge the changes. 
Moreover, the tool preserves the lines that remain unaffected by the changes from either the left or right versions, ensuring that they are in the final merged output. 

Finally, in the last step, \mt\ removes from the resulting file all the new lines and placeholders created before, generating its final result, which in this case is identical to the one yielded by a structured merge tool, shown before in Figure~\ref{fig:toString_structuredmerge}. 
In situations where conflicts arise during merging--- such as when both the left and right sides independently modify the condition of an \texttt{if} statement--- \mt\ employs the same process. However, in these conflicting scenarios, our tool reports the conflicting elements without the inclusion of the additional lines and placeholders introduced in its preprocessing phase. 



\section{Empirical Evaluation}


To assess \mt's potential for code merging, we compare it with semistructured and unstructured merge tools. 
The comparison with the first allows us to evaluate our separator-based technique in isolation, understanding how it advances previous results on semistructured merge.
The comparison with the latter provides insights into the benefits that our tool could bring to the standard practice.

In both cases, we evaluate the effectiveness of the merge tools using metrics that capture the tools' capacity to resolve merge conflicts while preserving the integrity and correctness of the merging process. 
This is essential to understand whether \mt\ can increase development productivity--- by reducing wasted time on resolving spurious conflicts--- without compromising software quality--- by letting actual conflicts escape. 
Our goal is to understand the strengths and weaknesses of \mt\ compared to other merge tools, so that software development teams can use that as input when deciding to adopt merge tools.

In particular, we observe metrics such as the number of merge conflicts and the number of files with conflicts.
This evidence, however, is not enough to justify adoption of a merge tool.
So we also investigate false positives and false negatives. 
The challenge associated with such a comparison criteria is establishing ground truth for integration conflicts, as this is not computable in this context~\cite{berzins1986merging, horwitz1989integrating}. 
Semantic approximations through static analysis or testing are imprecise and often expensive. Experts who understand the integrated code (possibly developers of each analyzed project) could determine truth, but not without the risk of misjudgement. 

As these options would imply into a reduced sample and limited precision guarantees, we prefer to \emph{relatively} compare the merge tools in pairs, regarding the occurrence of false positives and false negatives of one tool \emph{in addition} to the ones of the other. 
We do that by simply analyzing when two merge tools report different results for the same merge scenario --- a quadruple of commits, which we will call \emph{base}, \emph{left}, \emph{right}, and \emph{merge} commits, this last one being the final version of the merged code stored in the original project repository we analyze in our experiment. 
We identify files with merge conflicts reported by one tool but not by the other (\textit{added false positive (aFP)}), and files wiht merge conflicts reported by one tool but missed by the other (\textit{added false negative (aFN)}). 
Actual false positives and false negatives common to both tools are ignored, as we observe only situations in which the tools differ. 
This is aligned with our interest in relatively comparing the tools.

\subsection{Research Questions}

In summary, we evaluate \mt\ by addressing the following research questions. Further details are provided in Section~\ref{sec:methodology}.

\subsubsection{\textbf{RQ1}: Does \mt\ reduce the number of merge conflicts and files with merge conflicts compared to standard semistructured and unstructured merge tools?}

To answer this research question, we analyze the number of merge conflicts reported by each merge tool, and the files having these merge conflicts. 
This involves running the merge tools on the files associated with the base, left, and right commits of each merge scenario within our sample. 
Then, we count each conflict block reported by each tool, and we count the files in which these conflict blocks appear.
A conflict block reported by these tools is a set of lines enclosed by the standard conflict markers ``\verb|>>>>>>>|'' and ``\verb|<<<<<<<|'' as ilustrated before in Figures~\ref{fig:util_merge} and~\ref{fig:toString_structuredmerge}. 

\subsubsection{\textbf{RQ2}: Does \mt\ reduce the number of merge conflicts identification errors compared to standard semistructured and unstructured merge tools?} 

To address this research question, we count added false positives and false negatives as indicators of potential errors in merge conflict identification. 
It is important to emphasize that our focus here lies in comparing merge tools in a relative context rather than determining their absolute accuracy regarding a broad definition of conflict. Therefore, we do not need to assess the frequency of false positives and negatives when the merge tools exhibit identical behavior.

\subsection{Sampling}

We evaluate the merge tools with merge scenarios from the GitHub open-source Java projects listed in Table~\ref{tb:sample}. 

We selected these projects based on the following criteria:
\begin{itemize}
    \item Popularity, based on their number of stars on GitHub. All projects have at least 700 stars;
    \item Number of collaborators, with at least 50 collaborators. A larger number of collaborators might be an indicative of concurrent work, increasing the likewise of changes that need to be merged;
    \item Presence of merge commits on projects’ repository history. This is necessary so that we can derive merge scenarios to replicate the merge operation.
\end{itemize}


\begin{table}[htbp!]
    \centering
    \caption{Overview of the sample projects.}
    \begin{tabular}{lrr}
    \toprule
    \textbf{GitHub Repository} & \textbf{Stars} & \textbf{Contributors} \\
    \midrule
    apache/accumulo & 880 & 132 \\
    FasterXML/jackson-databind & 2.9k & 175 \\
    HubSpot/Singularity & 799 & 80 \\
    spring-projects/spring-boot & 56.3k & 827 \\
    mockito/mockito & 12k & 221 \\
    spring-projects/spring-framework & 43.6k & 544 \\
    libgdx/libgdx & 18.7k & 511 \\
    jenkinsci/jenkins & 17.6k & 659 \\
    alibaba/fastjson & 23.6k & 173 \\
    \bottomrule
    \end{tabular}
    \label{tb:sample}
\end{table}

To conduct our comparative analysis of these merge tools, we initially compile a dataset of merge commits from our sample projects excluding fast-forwarding merges, as it cannot cause merge conflicts. Each merge commit represents a merge operation within the project's history and serves as the basis for deriving a merge scenario. 
For each merge commit, we identify the corresponding merge scenario, which consists of a quadruple of commits. 
This quadruple comprises the merge commit itself, representing the official merge result in the project's repository, and the commits associated with the three versions involved in a three-way merge: the base commit, the left commit, and the right commit.
Given that each merge scenario contains four sets of files---one for each of the aforementioned commits --- we use the merge tools to merge the files within the base, left, and right commits. We then store the results of these merge operations for each merge scenario. 
As consequence of this step, we perform our study on 9510 sets of files from 1054 merge scenarios from our sample projects. 

\subsection{Methodology}
\label{sec:methodology}

To answer our research questions we gather evidence regarding the frequency of divergent results among the tools we compare, as well as instances of errors in conflict detection, namely added false positives and false negatives. 
We adopt the state-of-practice \texttt{diff3} as standard unstructured merge tool, the state-of-art \texttt{s3m}~\cite{cavalcantietal2017} as standard semistructured merge tool, and our solution \mt\ configured with the following Java language-specific syntactic separators: `\texttt{\{}', `\texttt{\}}', `\texttt{(}', `\texttt{)}' and `\texttt{;}'.

Following the merging process, we systematically collect data on the number of merge conflicts, the prevalence of conflicting files, added false positives, and added false negatives. 
A conflicting file is defined as a merged file containing at least one merge conflict. 
In cases where the merge results generated by the merge tools differ,
we compare these results with the merge result in the repository, and collect evidence for further analysis, particularly concerning false positives and false negatives. We consider merge tools to differ when (1) they report a different number of conflicts for the same file, or (2) the number of merge conflicts is the same, but the content of the files, ignoring whitespace (as it is semantically neutral in Java), is different.
In particular, when comparing two merge tools, $M$ and $N$, we compute an occurrence of added false positives to $M$ when a file has at least one merge conflict reported by $M$, while none are reported by $N$. 
Moreover, for these cases, $N$'s result must align with the outcome in the original project repository (obtained via the merge commit), indicating $N$'s successful and accurate resolution of merge conflicts in the file. 
Conversely, we compute an added false negatives to $M$ when it fails to report any merge conflicts in a file, yet the merged file deviates from the result in the repository merge commit. Furthermore, we require tool $N$ to report merge conflicts in the same file. 
In such instances, we consider that tool $M$ inaccurately resolved the merge conflicts, resulting in syntactically or semantically incorrect code compared to the expected outcome.

Finally, to consolidate our metrics regarding merge conflicts identification errors, we conducted a manual analysis of the potential false negatives. We compared the merge results with the merge commit to see whether differences were due to developer's manual changes, and whether there was any actual interference between the changes. If there are developer-made changes during the merge, then $M$ had integrated correctly, indicating a false positive from $N$ rather than a false negative from $M$.

We provide the scripts and data associated with this study in the online appendix.\footnote{Omitted for anonymity.}


\section{Results and Discussion}

Next, we present and discuss our findings, structured according to our the research questions.

\vspace{1.5mm}
 \noindent{\textbf{RQ1:} \textit{Does \mt\ reduce the number of merge conflicts and files with merge conflicts compared to standard semistructured and unstructured merge tools?}}
\vspace{1.5mm}

Table~\ref{tab:merge_conflicts_comparison} shows that embodying semistructured merge with language-specific syntactic separators, through the implementation of \mt, resulted in a reduction not only in merge conflicts, but also in files having these merge conflicts. This applies when we compare \mt\ to both the conventional unstructured merge \texttt{diff3} tool, and the semistructured merge \texttt{s3m} tool. Specifically, \mt\ reduced the number of merge conflicts by approximately 41\% and the number of conflicting files by about 40\% compared to \texttt{diff3}. Additionally, when comparing \mt\ to \texttt{s3m}, it can be observed that \mt\ managed to decrease the number of merge conflicts by around 13\% and the number of conflicting files by approximately 21\%. 

\begin{table}[h]
    \centering
    \begin{tabular}{lccc}
        \hline
        & \texttt{diff3} & \mt & \texttt{s3m} \\
        \hline
       \textbf{Merge conflicts} & 2413 & 1413 & 1632 \\
       \textbf{Conflicting files} & 1090 & 657 & 832 \\
        \hline
    \end{tabular}
    \caption{Comparison of merge conflicts and conflicting files.}
    \label{tab:merge_conflicts_comparison}
\end{table}

\begin{table*}[t]
    \centering
    \begin{tabular}{cccc}
        \hline
        & \texttt{diff3 vs.SESAME} & \texttt{diff3 vs.s3m} & \texttt{s3m vs.SESAME} \\
        \hline
       \textbf{Total} & 1145 & 1019 & 675 \\
       \textbf{(\%)*} & 12.04\% & 10.72\% & 7.10\% \\
        \hline
        \multicolumn{4}{l}{\small \footnotesize{* In relation to the total number of merged files.}} \\
    \end{tabular}
    \caption{How often the merge tools differ.}
    \label{tab:merge_conflicts_diff}
\end{table*}

Our solution is driven by the aspiration to leverage the benefits of a fully structured merge tool, while simultaneously minimizing the costs typically associated with such tools. We aimed to strike a balance between the advantages in merge conflicts detection of structured merge and the practical considerations of implementation complexity and resource requirements. By comparing our findings with prior research, we have strong indications that our objective has been met. In particular, our results closely mirror those documented by Apel et al.~\cite{apel2012structured} when they evaluated \texttt{JDime}, which is, to the best of our knowledge, the most representative structured tool for Java applications. They reported a 39\% reduction in merge conflicts when comparing their structured tool against traditional unstructured merge. This alignment between our findings and prior research not only validates the efficacy of our tool but also reinforces the potential benefits of leveraging structure to reduce merge conflicts during the merging process.

Finally, the \mt\ tool reduced the number of merge conflicts in our sample, but the specific merge conflicts reported by each merge tool are not necessarily identical. This discrepancy means that one tool might identify certain conflicts that another tool misses, and vice versa (as we will further discuss on the next research question), or a tool might report the same conflict in a different manner. Therefore, to gain a deeper understanding of the differences in the \mt's behavior, it is essential to analyze not just the overall number of conflicts each merge tool reports, but also the frequency and nature of the instances where the tools diverge. In that sense, Table~\ref{tab:merge_conflicts_diff} illustrates that when we textually compare the outputs produced by the merge tools, \mt\ diverges from \texttt{diff3} in approximately 12\% of the 9510 merged files in our sample. Additionally, \mt\ differs from \texttt{s3m} in about 7\% of these files. 

When inspecting the output produced by \mt, we observed that the tool generates several new lines, one for each language-specific syntactic separator. This results in situations where different alignments emerge when running unstructured merge on these files as part of the \mt\ process described in Section~\ref{sec:solution}. These problematic alignments occur because \mt\ identifies a greater match between files using the new, smaller lines created by the algorithm. This finer granularity introduced by the syntactic separators can sometimes lead to unexpected alignment issues, as the invoked unstructured merge tool may perceive the newly created lines as distinct changes. This phenomenon is a source of \mt's added false positives, as we further elaborate next in our discussion of the second research question.


\vspace{1.5mm}
 \noindent{\textbf{RQ2:} \textit{Does \mt\ reduce the number of merge conflicts identification errors compared to standard semistructured and unstructured merge tools?}}
\vspace{1.5mm}

Assessing the effectiveness of a merge tool involves more than just evaluating its merge conflict reduction capabilities; it also demands an analyzes of its conflict detection accuracy. Spurious merge conflict detection can significantly impede developers' productivity by leading them to spend valuable time resolving conflicts that are not genuine. Conversely, when a tool overlooks merge conflicts, it opens the door to potential syntactic and/or semantic errors in the source code, which can undermine the quality and stability of the software being developed. Therefore, a comprehensive evaluation of a merge tool should consider both its conflict reduction efficacy and its precision in identifying genuine conflicts. Thus, to answer our second research question, we relatively compare the merge tools in terms of added false positives and added false negatives as described in Section~\ref{sec:methodology}.

Starting with the false positives, Table~\ref{tab:fpfn_comparison} shows that \mt\ managed to reduce the number of added false positives in relation to both \texttt{diff3} (unstructured merge) and \texttt{s3m} (standard semistructured merge). In particular, \mt\ exhibited a significant reduction in the number of added false positives by approximately 91\% compared to \texttt{diff3} and by approximately 88\% compared to \texttt{s3m}. Note that for class-level declarations such as methods and fields, \mt\ and \texttt{s3m} behave the same way by definition. This is because both tools handle class-level merge conflicts using the same semistructured approach. Conversely, at the statement level, \texttt{s3m} and \texttt{diff3} exhibit the same behavior, as \texttt{s3m} relies on the traditional unstructured approach of \texttt{diff3}. However, \mt\ is enhanced by language-specific syntactic separators, which allows it to handle statement-level conflicts more effectively. This distinction explains the comparable improvements seen with \mt\ when compared to the other tools. Specifically, \mt\ reduces false positives from \texttt{diff3} at the class level by leveraging its semistructured approach, such as merge conflicts caused by the (re)ordering of methods and field declarations as illustrated before in Section~\ref{sec:motivation}. Additionally, \mt\ reduces false positives from both \texttt{diff3} and \texttt{s3m} at the statement level due to its enhanced handling of language-specific syntax thanks to the separators. This improvement is particularly evident in cases involving  changes to unrelated variable declarations or assignments on consecutive lines within a method declaration.

\begin{table}[b!]
    \centering
    \begin{tabular}{lcccc}
        \hline
        & \texttt{diff3} & \mt \\
        \hline
       \textbf{aFP} & 291 & 26 \\
       \textbf{aFN} & 0 & 168 \\
        \\
        \hline
        & \texttt{s3m} & \mt \\
        \hline
       \textbf{aFP} & 146 & 18 \\
       \textbf{aFN} & 1 & 48 \\
        \\
        \hline
        & \texttt{diff3} & \texttt{s3m}\\
        \hline
       \textbf{aFP} & 169 & 31 \\
       \textbf{aFN} & 4 & 124 \\
    \end{tabular}
    \caption{Comparison of merge conflicts identification errors in terms of false positives (aFP) and false negatives (aFN).}
    \label{tab:fpfn_comparison}
\end{table}

Nevertheless, \mt\ approach leads to its own false positives. The occurrence of these false conflicts can be attributed to the way \mt\ handles syntactic separators. By isolating these separators into new lines, \mt\ alters the structure of the code as demonstrated before in Figure~\ref{fig:toString_transformation}, which can sometimes lead to unexpected alignments. This misalignment occurs because \mt, when performing the unstructured merge as part of its process, becomes ``confused'' as it attempts to identify and align the modified text areas from the left and right versions. Consequently, unstructured merge misinterprets these changes, resulting in false conflicts that would not have been present if the separators were not isolated into separate lines.

Regarding false negatives, \mt\ follows a trend observed in prior studies~\cite{cavalcantietal2017,cavalcanti2019impact,tavares2019semistructured,seibt2021leveraging}: the more a merge tool leverages the structure of the code being merged, the higher the likelihood that some merge conflicts will go undetected. Note in Table~\ref{tab:fpfn_comparison} that both \mt\ and \texttt{s3m} show an increase in the number of added false negatives compared to \texttt{diff3}. Additionally, \mt, though to a lesser extent, also shows an increase in added false negatives compared to \texttt{s3m}. This phenomenon occurs because structured merge tools, while being more sophisticated in understanding and integrating changes based on the syntactic and semantic structure of the code, can sometimes overlook conflicts that simpler, unstructured tools would catch \textit{accidentally} as they threat everything as lines of text. By focusing on the low-level, finer granularity, organization of the code, the structure-driven tools may miss subtle conflicts among statements or expressions, leading to an increase in false negatives.

\begin{figure*}[t]
  \includegraphics[width=\linewidth]{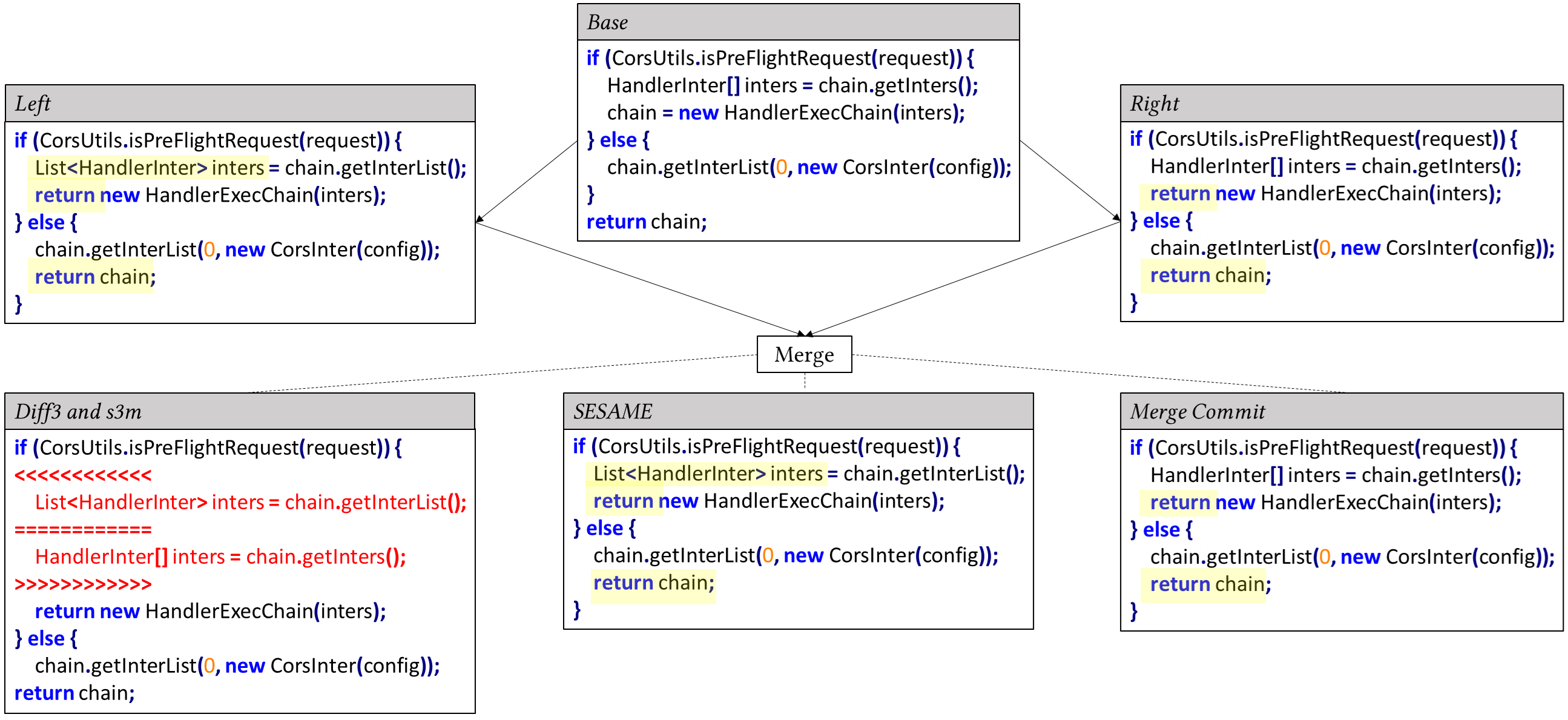}
  \caption{Merging \textit{AbstractHandlerMapping.java} from project SpringFramework (changes highlighted in yellow).}
  \label{fig:AbstractHandlerMapping_changes}
\end{figure*}

For instance,  consider the example taken from project SpringFramework in Figure~\ref{fig:AbstractHandlerMapping_changes}. In that scenario,  both developers moved the \texttt{return} statement to within an \texttt{if-else} statement. However, in the left version, the developer also changed the type of a variable from an array to a List. When merging these changes, both \texttt{diff3} and \texttt{s3m} report a merge conflict because part of the changes occurred on consecutive lines of code. In contrast, \mt\ does not report merge conflict because the ``;" in line 2 act as a separator of context. However, upon inspecting the original merged version in the merge commit (as illustrated in Figure~\ref{fig:AbstractHandlerMapping_changes} also), we observed that the output from \mt\ did not match the original version: the integrator chose to keep the array from the base version instead of the List from the left version. 

Besides that, the misalignment problem explained before, besides generating false conflicts, also causes \mt\ to sometimes fail to generate a conflict during the merging process. To understand how the misalignment leads to an added false negative, consider a base file that has the line ``\texttt{a().b(c).d();}'' with chained method calls. Suppose that one of the developers modified the argument of method \texttt{b}, resulting in ``\texttt{a().b(\textbf{e}).d();}''. Meanwhile, the other developer replaced the call to method \texttt{b} with a call to \texttt{g} as follows: ``\texttt{a()\textbf{.g(h(c))}.d();}''. When integrating these versions, \mt, using the parentheses as separators, ends up aligning the part ``\texttt{(c)}'' from the base with the one from right. By adding a new line before and after each separator, \mt\ interprets these lines as forming the largest alignment and, because of this, it accepts both modifications without conflict, resulting in ``\texttt{a().\textbf{g(h(e))}.d();}''. However, if there are semantic dependencies between the modifications, \mt\ would let merge conflict goes undetected.

Classifying such merge conflicts as a false positive or false negative is a delicate task because it requires understanding if and how the changes are related. At the statement and expression levels, this task might involve comprehending the behavior of the code and analyzing other files, which are beyond the scope of the studied tools. Consequently, a compromise exists: unstructured tools are overly conservative, treating everything uniformly as text, whereas structure-driven tools are more optimistic, assuming that changes to different code structure elements are safe to merge. This trade-off reflects the challenges inherent in balancing precision and comprehensiveness in merge conflict identification and resolution.

\subsection{Threats to Validity}

The generalizability of our results may be limited by the specific programming language and types of projects included in our evaluation. Our approach leverages Java language-specific syntactic separators, so its effectiveness may vary across different languages and coding styles. So, further studies are necessary to confirm the tool's applicability in other contexts. 

Besides that, the accuracy of our merge conflicts identification errors metrics, in particular false negatives, is a potential threat. There is the possibility that the developer modified the code during the merge process. So we cannot effectively identify false negatives. Therefore, a manual analysis was conducted for cases of potential false negatives. Given that this analysis was manual, there is a risk of incorrect judgment in evaluating these cases. However, we mitigated these risks by involving multiple authors in the process.

Finally, our sample represents only a subset of the actual code integration scenarios within the analyzed projects. Specifically, we only analyze scenarios that culminate in public repository merge commits. This excludes integrations performed using Git rebase or other Git commands that rewrite history, which do not result in merge commits and therefore are not captured in our analysis.

\section{Related Work}

The challenge of efficiently detecting and resolving merge conflicts in collaborative software development has led to significant research in the field of version control systems. 


\subsection{Semistructured merge}

Apel et al.~\cite{apel2011semistructured} proposed semistructured merge, a merge approach that \textit{partially} leverages the syntactic structure and static semantics of the underlying language, but without the performance overhead associated with fully structured merge. They implemented semistructured merge in the tool \texttt{FSTMerge} for C\#, Java, and Python. 
Tavares et al.~\cite{tavares2019semistructured} later extended \texttt{FSTMerge} to support JavaScript, finding that the benefits of semistructured merge might be limited for languages that support both commutative and non-commutative declarations at the same syntactic level, such as JavaScript.

Studies by Cavalcanti et al.\cite{cavalcantiassessing, cavalcantietal2017} provide evidence that semistructured merge can reduce the number of reported conflicts compared to traditional unstructured merge, though this benefit is not consistent across all projects and merge scenarios. Additionally, Cavalcanti et al.\cite{cavalcantietal2017} found that semistructured merge significantly reduces the number of false positives, but does not necessarily lead to fewer false negatives.
Cavalcanti et al.~\cite{cavalcanti2019impact} further compared semistructured merge with structured merge. They found that semistructured and structured merge differ in 24\% of the merge scenarios with merge conflicts, semistructured merge reports more false positives, whereas structured merge has more false negatives. In our work, we found that semistructured merge effectively reduces merge conflicts and false positives. However, it also exhibits a tendency towards false negatives, a tendency amplified by our approach that aims to mimic structured merge.

Seibt et al.~\cite{seibt2021leveraging}, in a comprehensive evaluation of various merge algorithms (ranging from unstructured to semistructured to structured, as well as combinations of them), found that combined strategies offer the best of both worlds. Their evidence suggests that these hybrid approaches can resolve as many merge conflicts as structured merge algorithms while achieving significantly lower runtimes per merge scenario. Our work follows this direction by further combining unstructured and semistructured in a merge tool designed to mimic structured merge.

\subsection{Structured merge}

Apel et al.~\cite{apel2012structured} introduced \texttt{JDime}, a state-of-the-art structured merge tool for Java applications. \texttt{JDime} dynamically switches between unstructured and structured merge depending on the presence of conflicts, optimizing the merging process. This work has inspired numerous recent advancements in the field. For instance, Lessenich et al.~\cite{LeBenich:2017} enhanced \texttt{JDime} by incorporating a syntax-specific lookahead to detect refactorings, renamings, and shifted code. Their approach significantly improved matching precision, achieving a 28\% increase while maintaining performance. 

Zhu et al.~\cite{Zhu:2019} built upon \texttt{JDime} with \texttt{AutoMerge}, which uses an adjustable \textit{quality function} to match nodes. This method maximizes the quality function to prevent the matching of logically unrelated nodes, thereby reducing false positive conflicts. Their findings indicate that \texttt{AutoMerge} successfully reduced the number of reported conflicts compared to the original \texttt{JDime}, albeit with a slight decrease in speed. Zhu et al.~\cite{zhu2022mastery} also proposed \texttt{Mastery}, a structured merge tool that utilizes both top-down and bottom-up traversal of abstract syntax trees (ASTs) to handle shifted code more effectively and elegantly compared to \texttt{JDime}, which relies solely on a top-down traversal restricted to level-wise AST node matching. Their approach demonstrates to be 2.4 times faster than \texttt{JDime}, and achieves a merge accuracy of over 80\% relative to the expected output of the merging process. 

Larsen et al.~\cite{larsen2022spork} proposed \texttt{Spork}, a structured merge tool tailored to the Java programming language, designed to preserve the formatting of merged code by reusing the source code from the input files during pretty-printing. \texttt{Spork} builds upon the merge algorithm of the 3DM merge tool for XML documents, as they found its core principles are applicable to Java. Their evaluation demonstrates that \texttt{Spork} significantly improves formatting preservation in over 90\% of the merged files compared to \texttt{JDime}, while also being 51\% faster. 

All of these works recognize the inaccuracy of unstructured merge, as well as the performance cost of structured tools. Here, we propose a tool that offers the benefits of structured tools without the inherent development complexity and performance cost. 


\subsection{Other kinds of tools}

Shen et al.~\cite{shen2019intellimerge} introduced \texttt{IntelliMerge}, a graph-based refactoring aware merging tool for Java programs. Their findings demonstrate that \texttt{IntelliMerge} reduces the number of merge conflicts by about 59\% compared to unstructured merge and by about 12\% compared to a semistructured merge tool. Additionally, \texttt{IntelliMerge} achieves a precision of around 88\% and a recall of 90\%, using the manually resolved version as the ground truth. Ellis et al.\cite{ellis2022operation} proposed \texttt{RefMerge}, an operation-based refactoring-aware merging tool. 
 Operation-based based tools record the types of code changes (operations) instead of simple textual changes. Built on top of Git, \texttt{RefMerge} supports 17 refactoring types, including Extract Method and Inline Method. 
Their research indicates that \texttt{RefMerge} can reduce the number of false positives while eliminating false negatives compared to \texttt{IntelliMerge}. However, they also noted that \texttt{RefMerge} can sometimes introduce new conflicts of its own. Our solution does not specifically target refactorings, so it can work in conjunction with these tools to enhance the accuracy of the merging process, as they have demonstrated that refactoring is a significant source of merge conflict identification errors.


Other works propose tools to address the consequences of merge tools' missed conflicts (false negatives) in the merging process. For instance, Souza et al.~\cite{Sousa2018} proposed \texttt{SafeMerge}, a semantic tool that checks whether a merged program does not introduce new unwanted behavior. They achieve that by combining lightweight dependence analysis for shared program fragments and precise relational reasoning for the modifications. 
They found that their proposed approach can identify behavioral issues in problematic merges that are generated by unstructured tools. Zhang et al. \cite{zhang2022using} introduced \texttt{GMERGE}, a tool that leverages k-shot learning with a large language model (GPT-3) to automatically suggest merge conflict resolutions based on the input merge conflict and merge histories. \texttt{GMERGE} achieved an overall accuracy of approximately 64\% in resolving semantic merge conflicts in Microsoft Edge, using the developer’s actual fixes as the ground truth. 
Towqir et al.~\cite{towqir2022detecting} proposed \texttt{Bucond}, a tool that models each version in a merge scenario as a graph and compares these graphs to extract entity-related edits, such as class renaming, to avoid post-merge source code build errors.
Their evaluation across three datasets demonstrated that \texttt{Bucond} detects conflicts with high precision and recall, although it missed some conflicts that were identified through automatic builds or manual inspection. We believe that our proposed tool, \mt, could be used in conjunction with these tools by applying them to the results generated by \mt. This approach has the potential to mitigate the observed tendency to let conflicts go undetected, which can result in build and semantic errors in the merged code.



\section{Conclusion} 

In this paper, we introduced a novel semistructured merge approach, implemented in the \mt\ tool, which leverages language-specific syntactic separators to better detect and resolve merge conflicts. Our approach aims to combine the strengths of structured merge tools, which utilize the syntactic structure of code, with the simplicity and performance of traditional unstructured tools. By incorporating language-specific syntactic separators, \mt\ mirrors the capabilities of fully structured merge tools. This allows it to more accurately detect and resolve merge conflicts, thereby reducing the number of false positives.

Our evaluation demonstrated that \mt\ improves conflict detection over standard semistructured and unstructured merge tools, achieving gains comparable to fully structured merge tools without incurring the same development and runtime overhead. However, we also identified a drawback shared with fully structured tools: a tendency to fail in detecting actual, true merge conflicts. Additionally, we observed that incorporating language-specific syntactic separators can sometimes cause textual alignment errors, leading to different types of merge conflicts identification errors. Future work will focus on further refining our approach, with a particular emphasis on addressing the textual alignment errors observed in the current implementation.



\bibliographystyle{ieeetr}
\bibliography{sample-base}


\end{document}